\documentclass[%
 reprint,
 superscriptaddress,
 amsmath,amssymb,
 aps,
 prl,
floatfix,
nolongbibliography
]{revtex4-2}

\usepackage{graphicx}% Include figure files
\usepackage{dcolumn}% Align table columns on decimal point
\usepackage{bm}% bold math
\usepackage{bbold}% bold math
\usepackage{xparse}% bold math
\DeclareDocumentCommand{\myfig}{m o}{Fig.~\ref{#1}\IfValueT{#2}{#2}}
\newcommand*\myeq[1]{Eq.~(\ref{#1})}

\newcommand*\kB{k_{\mathrm{B}}}
\newcommand*\ob{\omega_{b}}
\newcommand*\oR{\omega_{R}}
\newcommand*\ee{\mathrm{e}}
\newcommand*\rmd{\mathrm{d}}
\DeclareMathOperator{\erf}{erf}

\newcommand*\subscale{0.6}
\newcommand*\kTST{k_{\scalebox{\subscale}{$\mathrm{TST}$}}}
\newcommand*\kram{{\scalebox{\subscale}{$\mathrm{K}$}}}
\newcommand*\vu{{\scalebox{\subscale}{$\mathrm{V}$}}}

\newcommand*\kK{\kappa_\kram}
\newcommand*\phik{\phi_\kram}
\newcommand*\varphik{\varphi_\kram}
\newcommand*\kV{\kappa_\vu}

\newcommand*\varphiv{\varphi_\vu}
\newcommand*\tphiv{\tilde{\varphi}_\vu}

\newcommand*\splus{\scalebox{\subscale}{$+$}}
\newcommand*\sminus{\scalebox{\subscale}{$-$}}
\newcommand*\spm{\scalebox{\subscale}{$\pm$}}

\newcommand*\aplus{a_{\splus}}
\newcommand*\aminus{a_{\sminus}}

\newcommand*\gauss{\varrho}
\newcommand*\gaussq{\rho}
\newcommand*\gpos{\gaussq^{\splus}}
\newcommand*\gneg{\gaussq^{\sminus}}
\newcommand*\Gauss{\mathbb{F}}

\newcommand*\jpos{j^{\splus}}

\newcommand*\amsmall{a_{\scalebox{0.4}{$-$}}}

\newcommand*\taus{\tau_\mathrm{s}}
\newcommand*\taur{\tau_\mathrm{r}}
\newcommand*\tstop{t_\mathrm{h}}
\newcommand*\teq{\!=\!}
\newcommand*\tgt{\!>\!}
\newcommand*\tlt{\!<\!}

\begin{document}

\title{On the classical reaction rate and the first-time problems of Brownian motion}

\author{Aihua Zhang}
\email{zah7903@gmail.com}
\affiliation
{Global AI Drug Discovery Center, College of Pharmacy and Graduate School of Pharmaceutical Sciences, Ewha Womans University, Seoul 03760, Republic of Korea}

\author{Sun Choi}
\email{sunchoi@ewha.ac.kr}
\affiliation
{Global AI Drug Discovery Center, College of Pharmacy and Graduate School of Pharmaceutical Sciences, Ewha Womans University, Seoul 03760, Republic of Korea}

\date{\today}

\begin{abstract}
We have developed efficient techniques to solve the first-time problems of Brownian motion.
Based on a time-scale separation of recrossings, we show that Eyring's transmission coefficient ($\kappa$) equals to the one ($\kV$) corresponding to an absorbing boundary consistent with the transition state theory, which is greater than the one ($\kK$) derived by Kramers.
We also propose methods for reaction rate determination by analyzing short-time trajectories from the barrier maximum, and discuss the relation to the reactive flux method and the significance of reaction coordinates.
\end{abstract}
%\keywords{Suggested keywords}%Use showkeys class option if keyword
                              %display desired
\maketitle

In 1940, Kramers published a seminal paper~\cite{Kramers1940} in which he derived the expressions of transmission coefficients ($\kK$) of thermally activated barrier crossing from a Brownian motion model of chemical reactions.
The development in the following fifty years had been thoroughly reviewed by H\"{a}nggi et al.~\cite{Hanggi1990} and a brief history of reaction rate theory was presented by Pollak and Talkner~\cite{Pollak2005} a century after Einstein's work on Brownian motion~\cite{Einstein1905}.
Kramers' derivation was performed separately for the cases of very weak friction and moderate-to-strong damping.
In this work, we focus on the latter case and show that $\kK$ corresponds to the survival probability ($\lim_{t\rightarrow \infty}\Phi_\kram(t)$) for an ensemble of systems having crossed the barrier maximum never to recross a specific absorbing boundary involving both the reaction coordinate ($q$) and its conjugate momentum ($p$).
We then work on a transmission coefficient ($\kV$) for an absorbing boundary specified only in $q$ and consistent with the transition state theory (TST)~\cite{Wigner1938}, which is intimately related to the recurrence time problem of Brownian motion mentioned by Wang and Uhlenbeck~\cite{Uhlenbeck1945} or the distribution of zero-crossing intervals of random processes referred to in the mathematical literature~\cite{Blake1973}.
We develop efficient techniques to determine $\kV$ and related $\Phi_\vu(t)$.
Based on a time-scale separation of recrossings, we find that Eyring's transmission coefficient ($\kappa$)~\cite{Eyring1935} can be interpreted as $\kV$, which is greater than $\kK$.
In the end, we propose methods to calculate reaction rates by short-time simulations starting from the barrier maximum, and discuss the relation to the reactive flux method and the importance of reaction coordinates.
In appendix, we demonstrate that the technique can also be applied to the first passage time (FPT) problem of Brownian motion.

\begin{figure}
\includegraphics[scale=1]{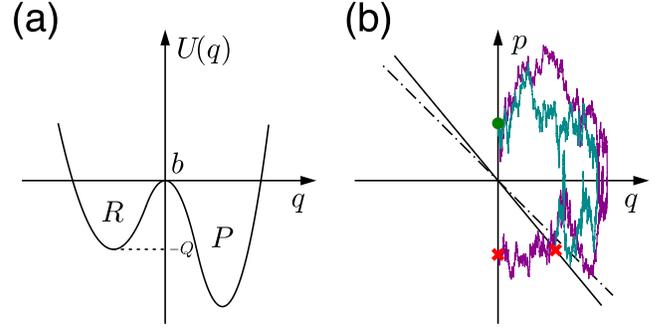}
\caption{\label{fig1} (a) A schematic graph of the PMF of an elementary reaction. (b) Example random trajectories that stop at the absorbing boundary K (cyan) and V (purple). 
%The dot-dashed line corresponds to $p=q$.
}
\end{figure}

\begin{figure*}
\includegraphics[scale=1.]{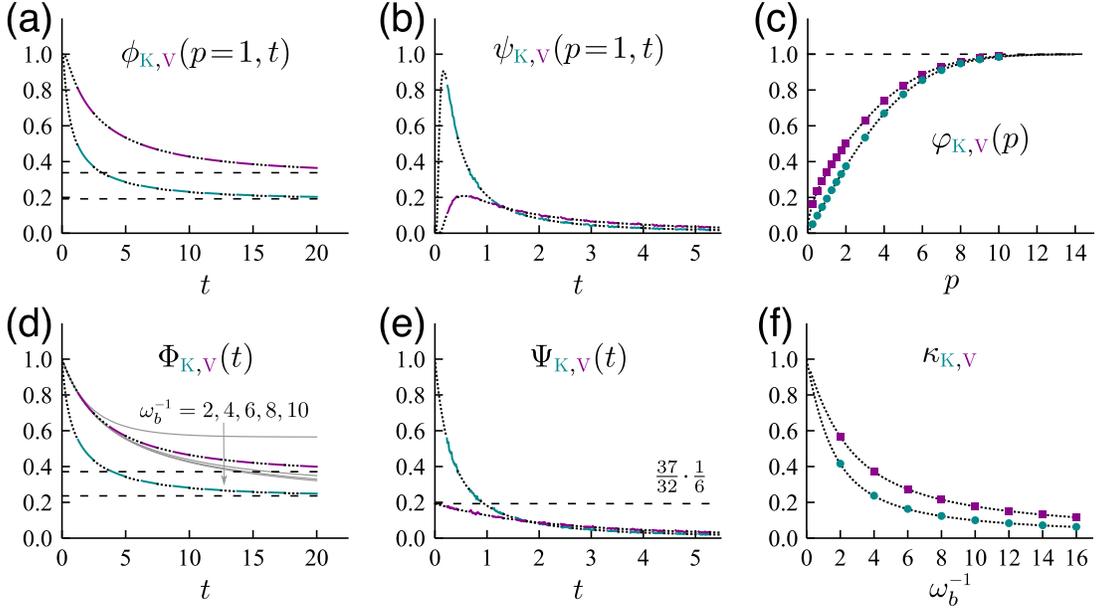}
\caption{\label{fig2} (a) $p$-resolved survival probabilities up to time $t$. (b) $p$-resolved probability densities of recrossing at time $t$. (c) $p$-resolved probabilities of never-recrossing. (d) $\jpos$-averaged survival probabilities up to time $t$. (e) $\jpos$-averaged probability densities of recrossing at time $t$. (f) Transmission coefficients. The solid (dotted) curves or scattered symbols correspond to simulated (calculated) data. The cyan (purple) curves are for the boundary K (V). Curves in (a-e) are generated with $\ob^{-1}=4$. $\Phi_\vu$-curves for other $\ob^{-1}$ values are shown as gray in (d). The calculated and simulated curves are plotted alternatively for clarity.}
\end{figure*}

Following Kramers' treatment, we start with the familiar Langevin equation for Brownian motion in a force field, i.e.
\begin{equation}
    m \ddot{q} +\gamma \dot{q} + U'(q) = \xi(t),
    \label{langevin}
\end{equation}
where each overdot denotes a derivative with respect to time, $m$ the effective mass, $\gamma$ the friction coefficient and $U(q)$ the potential of \emph{mean} force (PMF) for an elementary reaction along the reaction coordinate.
We assume a canonical ensemble and 
$\xi(t)$ is Gaussian white noise of zero mean with the correlation function being
%\[ \left<\xi(t)\right>=0 \]
\[ \left<\xi(t)\xi(s)\right>=2\gamma \kB T \delta(t-s), \]
where $\left<\cdot\right>$ stands for canonical ensemble average, $\kB$ the Boltzmann constant, $T$ the temperature and $\delta(\cdot)$ the Dirac function.
We take the reaction coordinate as a \emph{predefined} dynamical variable coupled to some measuring process, by which locally stable reactant ($q\tlt 0$) and product ($q\tgt 0$) can be distinguished.
%The dividing surface ($q\mkern-2mu =\mkern-2mu 0$) is naturally chosen at the barrier maximum of $U(q)$, i.e.\ the least probable configuration~\cite{Evans1938}.
The dividing surface ($q\teq 0$) is naturally chosen at the barrier maximum of $U(q)$, i.e.\ the least probable configuration~\cite{Evans1938}.
A schematic graph of $U(q)$ is shown in \myfig{fig1}{(a)}.
In the following, we will change units so that
$ \gamma = m = \kB T = 1 $
and especially, the time unit will be $m/\gamma$.
We assume that around the barrier maximum and the reactant minimum ($q\teq -L$), $U(q)$ can be approximated as 
\begin{equation}
U(q) = \begin{cases}
-\frac{1}{2} \ob^2 q^2 & q\sim b \\
\frac{1}{2} \oR^2 (q+L)^2-Q &q\sim R,
\end{cases}
\label{pot-of-q}
\end{equation}
where $Q$ is the barrier height.
With the system in thermal equilibrium, the forward flux ($R\!\rightarrow\! P$) through the dividing surface and the reactant population can be calculated by 
$ \jpos = \left<p\,\theta(p) \delta(q) \right> $
and
$n_R = \left< \theta(-q) \right> $ ($\theta(\cdot)$ is the Heaviside function), respectively.
By the flux-over-population method~\cite{Farkas1927}, the classical TST rate is obtained as
\begin{equation}
\kTST = \frac{\jpos}{n_R} \approx \frac{\oR}{2\pi}\ee^{-Q}.
\label{flux-over-pop}
\end{equation}
However, $\kTST$ overestimates the actual reaction rate ($k$) and an \textit{ad hoc} fudge factor ($\kappa$) was introduced by Eyring to account for the discrepancy, which was attributed to the system's recrossing the dividing surface, i.e.
\begin{equation}
    k = \kappa \cdot \kTST.
    \label{kappa-def}
\end{equation}
Since recrossing eventually happens in an equilibrium, this hints at a problem of time-scale separation. 

%\[ \rho(p,q) = Z^{-1} \ee^{-\frac{p^2}{2}-U(q)} \]
%\[ j = \int_0^\infty\!\rmd p\, p\,\rho(p,0)  = Z^{-1} \]
%\begin{eqnarray*}
%n_R &\approx& Z^{-1} \int_{-\infty}^{\infty}\!\rmd p\,\ee^{-\frac{p^2}{2}}
%\int_{-\infty}^{\infty}\!\rmd q\,\ee^{-\frac{1}{2}\oR^2(q+L)^2+Q} \\
%&=& Z^{-1}\frac{2\pi}{\oR}\ee^Q
%\end{eqnarray*}

%Define $\Delta \equiv \sqrt{1+4\ob^2}$ and $a_{\scalebox{0.6}{$\pm$}} \equiv (\Delta \pm 1)/2$ for convenience. 

%Kramers derived a steady state diffusion solution of
%\[
%    \rho(p,q)=Z^{-1} \frac{1+\varphik(p-\aplus q)}{2}\cdot \ee^{-\frac{p^2}{2}+\frac{\ob^2}{2}q^2}\quad q\sim b
%\]

%\[
%    \varphik(u)=\erf(\sqrt{\aminus/2}\,u)
%\]

%\begin{eqnarray*}
%    j &=& \int_{-\infty}^\infty\!\rmd p\, p\, \rho(p,0) \\
%      &=& Z^{-1} \int_{-\infty}^\infty\!\rmd p\,
%          \frac{1+\varphik(p)}{2}\cdot p\, \ee^{-\frac{p^2}{2}} \\
%      &=& Z^{-1} \int_{0}^\infty\!\rmd p\,
%          \varphik(p)\cdot p\, \ee^{-\frac{p^2}{2}} \\
%      &=& Z^{-1} \kK
%\end{eqnarray*}

%\begin{eqnarray}
%    m \ddot{q} &=& F_\mathrm{friction} + F_\mathrm{mean} + F_\mathrm{random}\\
%    &=& -\gamma \dot{q} - U'(q) + \xi(t)
%\end{eqnarray}
%\[ \kK = \sqrt{\frac{\Delta-1}{\Delta+1}} \]

If we define $a_{\spm}\!\equiv\!(\sqrt{1+4\ob^2} \pm 1)/2$, then Kramers' transmission coefficient has the form of
$\kK\teq \sqrt{{\aminus}/{\aplus}}$.
The fact that $\kK$ depends only on $\ob$ implies that it is related to dynamics around the barrier maximum, for which the Langevin equation is reduced to
\begin{equation}
    \ddot{q} + \dot{q} - \ob^2 q = \xi(t).
    \label{le-bar-top}
\end{equation}
The Kolmogorov backward equation~\cite{Pontryagin1989} corresponding to \myeq{le-bar-top} reads
\begin{equation}
    \phi_t = (\ob^2 q - p) \phi_p + p \phi_q + \phi_{pp},
\end{equation}
where each subscript denotes a partial derivative, and $\phi(q,p,t)$ is the probability that a representative point initially situated at $(q,p)$ survives from hitting an absorbing boundary for the time up to $t$.
If $\phi=\phi(u,t)$ and $u = p + \aplus q$, the problem is further simplified to
\begin{equation}
    \phi_t = \aminus u \phi_u + \phi_{uu},
\end{equation}
which can be solved by 
\begin{equation}
    \phik(u,t)=\erf(\sqrt{\aminus(1+\coth(\aminus t))}\,u/2)
\end{equation}
for the initial condition, $\phi(u,0)=\theta(u)$, and the absorbing boundary condition, $\phi(0,t)=0$.
As a result, with an absorbing boundary set up at $p + \aplus q = 0$ (K), the probability of no-recrossing up to $t$ for a system crossing the dividing surface with momentum $p$ will be $\phi_\kram(p,t)$, i.e.
\begin{eqnarray}
    \phi_\kram(p,t)=\mathrm{Pr}\{\,& &p(\tau)+\aplus q(\tau)>0\ \mathrm{for}\ 0<\tau\leqslant t\ | \nonumber\\  & &q(0)=0,\ p(0)=p\,\}.
    \label{phi-def}
\end{eqnarray}
An example random trajectory in the phase space that starts from $(0,1)$ and ends at the boundary K is shown in \myfig{fig1}{(b)}.
Since the inverted harmonic potential in \myeq{le-bar-top} extends to negative infinity, there is a finite probability of never-recrossing, i.e.\ $\varphik(p)\equiv \lim_{t\rightarrow\infty}\phi_\kram(p,t)>0$.
The average survival probability for the ensemble of the equilibrium forward flux $\jpos$ can be calculated by $\Phi_\kram(t) = \left<  \phi_\kram(p,t) \right>_{\jpos} \equiv \left< \phi_\kram(p,t) p\,\theta(p)\delta(q) \right>$, which gives the result of
\begin{equation}
    \Phi_\kram(t) = \sqrt{\frac{\aminus}{\aplus-\ee^{-2\amsmall t}}}.
\end{equation}
It is easy to see $\kK = \Phi_\kram(\infty)$, by which $\kK$ can be interpreted as the fraction of $\jpos$ that will never recross the boundary K on the infinite inverted harmonic potential.
The $p$-resolved and $\jpos$-averaged probability density of recrossing time can be easily obtained by $\psi_\kram(p,t)=-\dot{\phi}_\kram(p,t)$ and $\Psi_\kram(t)=-\dot{\Phi}_\kram(t)$, respectively.
Curves of $\phi_\kram$, $\psi_\kram$, $\varphi_\kram$, $\Phi_\kram$ and $\Psi_\kram$ calculated for $\ob^{-1}\teq 4$ are shown in \myfig{fig2}{(a-e)}, and $\kK$ as a function of $\ob^{-1}$ is presented in \myfig{fig2}{(f)}.
All calculated data agree with the results analyzed according to \myeq{phi-def} from trajectories simulated by integrating \myeq{le-bar-top} using the BAOAB algorithm~\cite{Leimkuhler2013} with a time step of $10^{-4}$.

%For initial condition, $\varphi(u,0)=\delta(u-u_0)$, and absorption boundary, $\varphi(0,t)=0$:
%\[
%    \varphi(u,t;u_0)=\phi(u,t;u_0) - \phi(u,t;-u_0)
%\]
%\[
%    \phi(u,t;u_0) = \sqrt{\frac{\aminus}{2\pi (1-\ee^{\uminus 2\amsmall t})}}\, 
%    \exp\!\left[{-\aminus t - \frac{\aminus}{2}\frac{(u - u_0 \ee^{\uminus\amsmall t})^2}{1-\ee^{\uminus 2\amsmall t}}}\right]
%\]

%\begin{eqnarray}
%    \psi_\kram(p,t) &=& -\dot{\phi}_\kram(p,t) \nonumber\\
%    &=& \sqrt{\frac{2\aminus^3}{\pi}}\, \frac{p\,\ee^{-2\amsmall t - \frac{1}{4}\amsmall (1+\coth(\amsmall t)) p^2}}{\sqrt{(1-\ee^{-2\amsmall t})^{3}}}
%\end{eqnarray}

%\begin{eqnarray}
%    \Phi_\kram(t) &=& \int_0^\infty\!\rmd p\,\phik(p,t)\cdot p\,\ee^{-\frac{p^2}{2}} \nonumber\\
%    &=& \sqrt{\frac{\aminus}{\aplus-\ee^{-2\amsmall t}}}
%\end{eqnarray}

%\begin{eqnarray}
%    \Psi_\kram(t) &=& -\dot{\Phi}_\kram(t) \nonumber\\
%    &=& \ee^{-2\amsmall t} \left(\frac{\aminus}{\aplus-\ee^{-2\amsmall t}} \right)^{3/2}
%\end{eqnarray}

We proceed to solve the same problem but with a perpendicular absorbing boundary set up at $q\teq 0$ (V), which is consistent with the dividing surface for the transition state theory.
An example random trajectory that starts from $(0,1)$ and recrosses $q\teq 0$ with $p\tlt 0$ is shown in \myfig{fig1}{(b)}.
A fundamental property of the random trajectories in the phase space is that any recrossing through $q=q_0$ takes place by winding around the point of $(q_0,0)$~\cite{McKean1962}.
If a system starts from $(q_0,p_0)$, evolves according to \myeq{le-bar-top} for time $t$ and arrives at $(q,p)$, there exists an analytic expression~\cite{Risken1996} for the transition probability density (TPD), $\gauss(q,p,t;q_0,p_0)$.
Suppose a system starts from $q\teq 0$ with $p\tgt 0$ or $p_0\tgt 0$.
Let $\varphiv(p)$ denote the probability of no-recrossing,
$\tphiv(p_0)$ the probability of eventually staying in $q\tgt 0$, i.e.
\[
    \tphiv(p_0) = \lim_{t\rightarrow \infty}
    \int_{-\infty}^\infty\!\rmd p
    \int_0^\infty\!\rmd q\, \gauss(q,p,t;0,p_0),
\]
and $\Gauss(p_0,p)\rmd p$ the probability of winding back through $(p,p+\rmd p)$, i.e.
\[
    \Gauss(p_0,p) = \int_0^\infty\!\rmd t\, p\,\gauss(0,p,t;0,p_0).
\]
Then we can decompose $\tphiv(p)$ into cases of no-recrossing and even-times recrossing as
\begin{equation}
    \tphiv(p_0) = \varphiv(p_0) + \int_0^\infty\!\rmd p\, \Gauss(p_0,p)\,\varphiv(p).
    \label{varphiv-decomp}
\end{equation}
By discretizing the above equation and solving the resulted system of linear equations, we obtain $\varphiv(p)$ and determine the transmission coefficient for the boundary V by $\kV =  \left< \varphiv(p) \right>_{\jpos}$.

\begin{figure}
\includegraphics[scale=1.]{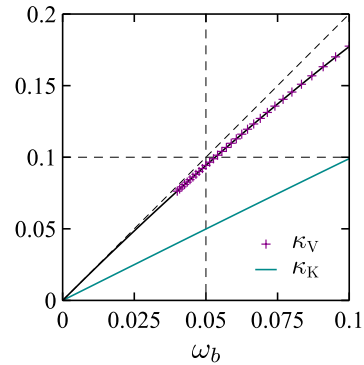}
\caption{\label{fig-diff-limit} Asymptotic behaviours of $\kV$ and $\kK$ at the diffusive limit ($\ob\!\rightarrow\!0$).
Numerical values of $\kV$ can be well fitted with $1.998\,\ob\!-\!2.249\,\ob^2$ (the solid black line).}
\end{figure}

We can also use the fundamental winding property to derive the recrossing time distributions up to a finite time.
We assume $p\tgt 0$ and $p_0\tgt 0$ as before.
First, we simplify the notations for the same-place TPD at $q\teq 0$ by defining $\gaussq^{\spm}(p,t;p_0) \equiv \gauss(0,\pm p,t;0,p_0)$.
Then we construct a winding operator ($\odot$) that acts on two successive TPDs of $\gaussq^{\mathrm{i}}$ and $\gaussq^{\mathrm{ii}}$, and produces the compounded TPD:
\begin{equation}
    \gaussq^{\mathrm{ii}} \odot \gaussq^{\mathrm{i}}
    \equiv \int_0^\infty\!\rmd p'\! \int_0^t\!\rmd t'\, \gaussq^{\mathrm{ii}}(p,t-t';p')\ p' \gaussq^{\mathrm{i}}(p',t';p_0).
    \label{winding-op}
\end{equation}
The winding operator is associative but not commutative, and the involved time convolution can be numerically performed using a fast Fourier transform algorithm.
Let $\zeta_1$ be similarly defined as $\gneg$ but for the first-time arrival.
Since $\gneg$ and $\gpos$ represent the odd-times and even-times arrivals, respectively, we have a relation of
\begin{equation}
    \zeta_1 = \gneg - \zeta_1 \odot \gpos.
    \label{zeta1-winding}
\end{equation}
By substituting the above relation into itself recursively, we obtain a series representation of $\zeta_1$ as
\begin{equation}
    \zeta_1 = \gneg - \gneg \gpos + \gneg {\gpos}^2 - \gneg {\gpos}^3 + \cdots,
    \label{zeta1-series}
\end{equation}
where the winding operator is implied between neighboring TPDs.
By discretizing $\gaussq^{\spm}$ and performing a certain number ($n$) of iterations, $2^n$ terms of \myeq{zeta1-series} can be accumulated until achieving a converged $\zeta_1$, from which the survival probability of $\phi_\vu(p,t)$ similarly defined as in \myeq{phi-def} can be derived as
\begin{equation}
    \phi_\vu(p,t) = 1-\int_0^t\!\rmd t'\! \int_0^\infty\!\rmd p'\, p'\zeta_1(p',t';p).
\end{equation}
Other functions of $\psi_\vu$, $\Phi_\vu$ and $\Psi_\vu$ follow similarly from $\phi_\vu$ as their K counterparts, and are shown together with $\varphiv$ and $\kV$ in \myfig{fig2}, which manifests that all calculated curves for the boundary V also agree with the simulated ones.
We can see $\kV\!>\!\kK$ in \myfig{fig2}{(f)}, which is consistent with the relative position of K and V boundaries.
It is worthy to note that $\Psi_\vu(0^{\splus})$ equals to the product of the magic number ($37/32$) derived by Wong~\cite{Wong1966,Wong1970} and the cubic Taylor coefficient ($1/6$) of the covariance function for \myeq{le-bar-top} (see \myfig{fig2}{(e)}).
The fact of $\Psi_\vu(0^{\splus})$ being constant can be related to 
the asymptotic form of $\gaussq^{\spm}$ at $t\!\rightarrow\!0$.
% Mar 27, 2023, on reviewer's suggestion
It is well known that $\kK$ is asymptotic to $\ob$ at the diffusive limit (see Eq.~(17) in Ref.~\cite{Kramers1940}).
We conjecture that $\kV\!\sim\!2\ob$ as $\ob\!\rightarrow\!0$ by numerical calculation (see \myfig{fig-diff-limit}).

%Let $\zeta_n(p,t;p_0)\,\rmd p\,\rmd q$ be the probability at $(0,(-1)^n p)$ for the $n^\mathrm{th}$ time at time $t$ starting from $(0,p_0)$.

%\[
%    \gauss_0^{\splus}(p,t;p_0) = \gauss(p>0,q=0,t;p_0)
%\]
%\[
%    \gauss_0^{\sminus}(p,t;p_0) = \gauss(-p<0,q=0,t;p_0)
%\]

%\begin{eqnarray}
%    \Omega[\zeta_n] &=& \zeta_{n+1}(p,t;p_0) \nonumber\\ 
%    &=&\int_0^\infty\!\rmd p'\!
%    \int_0^t\!\rmd t'\,\zeta_1(p,t-t';p')\cdot p'\zeta_n(p',t';p_0)
%\end{eqnarray}

%\begin{equation}
%    \sum_{n=0}^\infty \Omega^n[\zeta_1] = \gauss_0^{\splus} + \gauss_0^{\sminus}
%\end{equation}

%\begin{equation}
%    \psi_\vu(p,t) = \int_0^\infty\!\rmd p'\, p'\zeta_1(p',t;p)
%\end{equation}

%\begin{equation}
%    \sum_{n=0}^\infty \Omega^{2n}[\zeta_1] = \gauss_0^{\sminus}
%\end{equation}

%\begin{widetext}
%\begin{eqnarray}
%    \zeta_1(p,t;p_0) +
%    \int_0^\infty\!\rmd p_2\!
%    \int_0^\infty\!\rmd p_1\!
%    \int_0^t\!\rmd t_2\,
%    \int_0^{t_2}\!\rmd t_1\,
%    \zeta_1(p,t-t_2;p_2) \cdot p_2
%    \zeta_1(p_2,t_2-t_1;p_1) \cdot p_1\zeta_1(p_1,t_1;p_0)
%\end{eqnarray}
%\end{widetext}

\begin{figure*}
\includegraphics[scale=1]{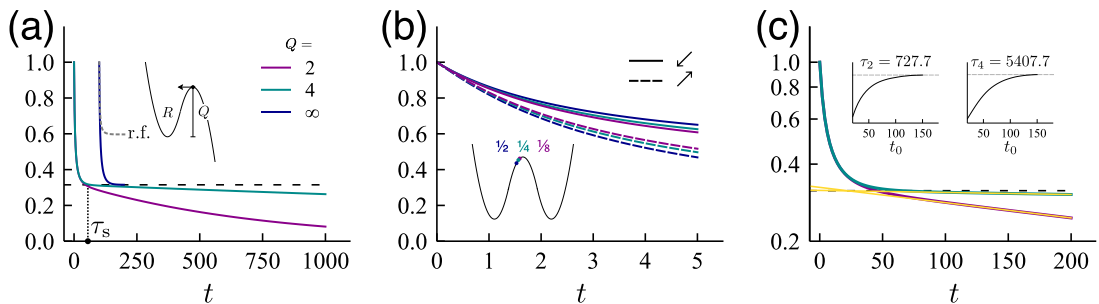}
\caption{\label{fig3} (a) $\Phi_\vu$-curves for the backward flux on the inset potential with different barrier heights. The curve labelled by r.f.\ is calculated by the reactive flux method. The curves for $Q\teq \infty$ are shifted to right for clarity. (b) The uphill ($\nearrow$) and downhill ($\swarrow$) $\Phi_\vu$-curves for different deviations in $\kB T$ away from the barrier maximum of the inset potential. (c) The logarithmic plots of $\Phi_\vu$-curves for $Q\teq 2,4$ in (a) up to a stop time of $\tstop\teq 200$. The insets show the variation of exponential decaying time constants obtained by linear fitting for the interval of $[t_0,\tstop]$. The yellow lines correspond to the asymptotic values.
}
\end{figure*}

So far we have studied the dynamics on an inverted harmonic potential, which approximates a barrier maximum with $Q\!=\!\infty$.
We continue to investigate the dynamics both on a barrier maximum and in a reactant minimum.
We model the related $U(q)$ using a differentiable piecewise-quadratic potential (see the inset of \myfig{fig3}{(a)}), which is parameterized by $\ob$, $\oR$ and $Q$ ($<\!\infty$).
By the flux-over-population formalism for $\kTST$, the transmission coefficient can be understood as a fraction of the forward flux that actually turns into products.
On the other hand, $\kTST$ is equivalent to the reciprocal of average time that a system in equilibrium spends in the reactant region ($q<0$).
So we can similarly calculate $\Phi_\vu$ and $\Psi_\vu$ from trajectories for a \emph{backward} flux ensemble, and then $\kTST$ can be determined as
\begin{equation}
    \kTST^{-1} = \int_0^\infty\!\rmd t\, t\,\Psi_\vu(t)
          = \int_0^\infty\!\rmd t\, \Phi_\vu(t).
          \label{ktst-inv}
\end{equation}
Two example $\Phi_\vu$-curves with $\ob^{-1}\!=\!\oR^{-1}\!=\!5$ and $Q\!=\!2,4$ are shown in \myfig{fig3}{(a)}, from which we observe a time-scale separation of recrossings.
A $1\!-\!\Phi_\vu(\taus)$ fraction of systems recross in a short time of $t\!<\!\taus$, while the remaining $\Phi_\vu(\taus)$ fraction of systems are supposed to traverse the minimum and recross in a slowly decaying way.
If only the long-time part ($t\!>\!\taus$) is resolved, the corresponding rate will be
\begin{equation}
    k^{-1} = \Phi_\vu(\taus)^{-1}\int_{\taus}^\infty\!\rmd t\, t\,\Psi_\vu(t).
    \label{k-inv}
\end{equation}
If we define $\epsilon=\int_0^{\taus}\!\rmd t\, (\Phi_\vu(t) - \Phi_\vu(\taus))\,/\int_0^\infty\!\rmd t\, \Phi_\vu(t)$, two rates are related by
\begin{equation}
    k = (1-\epsilon)^{-1}\,\Phi_\vu(\taus)\,\kTST.
\end{equation}
By comparing $\Phi_\vu$-curves of finite $Q$ with the $Q\!=\!\infty$ case (see \myfig{fig3}{(a)}), we find that the short-time part can be attributed to the dynamics on the barrier maximum, and it is a good approximation to take $\Phi_\vu(\taus)\approx \kV$ even for a small barrier height of $Q\!=\!2$.
If $\epsilon\!\ll\!1$ is also the case as in typical conditions, we obtain $k \approx \kV \cdot \kTST$, which establishes an interpretation of $\kappa$ in \myeq{kappa-def} as $\kV$.

%\begin{eqnarray*}
%    \epsilon &=& \int_0^{\taus}\!\rmd t\, (\Phi_\vu(t) - \Phi_\vu(\taus)) \\
%    \kTST^{-1} &=& \int_0^{\taus}\!\rmd t\, \Phi_\vu(t) + \int_{\taus}^\infty\!\rmd t\, \Phi_\vu(t) \\
%    \Phi_\vu(\taus) k^{-1} &=&  \int_{\taus}^\infty\!\rmd t\, t \Psi_\vu(t) \\
%    &=& -\int_{\taus}^\infty\!t \rmd \Phi_\vu(t) \\
%    &=& -\left[t \Phi_\vu(t)\vert_{\taus}^{\infty}- \int_{\taus}^\infty\!\rmd t \Phi_\vu(t) \right]\\
%    &=& \taus \Phi_\vu(\taus) + \int_{\taus}^\infty\!\rmd t \Phi_\vu(t)\\
%    &=& \kTST^{-1} - \epsilon \\
%    \Phi_\vu(\taus) \kTST &=& (1 - \epsilon / \kTST^{-1}) k \\ 
%\end{eqnarray*}

%\begin{equation}
%\kTST = \left< \tr(-p) \right>_{\jpos}^{-1}
%      \equiv \left< \tr(-p) p\,\theta(p)\delta(q) \right>^{-1}
%\end{equation}

The above elaboration forms the basis for calculating the reaction rate from simulations started from a barrier maximum.
Given a reaction coordinate, the barrier maximum can be determined by symmetry between $\Phi_\vu$-curves for the forward and backward flux.
As shown in \myfig{fig3}{(b)}, a slight deviation of $\kB T/8$ from the barrier maximum will give rise to apparent difference between the uphill and downhill $\Phi_\vu$-curves.
If $\kTST$ is available, we only need to simulate $\Phi_\vu(t)$ up to a short stop time ($\tstop$) less than $\taus$.
By first scaling the time unit so that $-\dot{\Phi}_\vu(0^{\splus})$ matches with $37/192$ and then finding $\ob^{-1}$ by comparing the resulted curve with $\Phi_\vu$-curves in \myfig{fig2}{(d)}, we can determine the reaction rate as $\kV \kTST$ after reading out $\kV$ from \myfig{fig2}{(f)}, and get an estimation of $\taus$ at the same time.
This approach resembles the reactive flux (r.f.) method~\cite{Yamamoto1960,Chandler1978}, but the first-time problem is not dealt with in the reactive flux method, and as shown in \myfig{fig3}{(a)}, the corresponding $\Phi_\mathrm{r.f.}(t)=\left< \theta(q(t)) \right>_{\jpos}$ overcounts the probability by even-times recrossings.

On the other hand, if the long-time recrossing dynamics obeys the phenomenological exponential decaying~\cite{Chandler1978}, i.e.\ $\Phi_\vu(t)\!=\!\Phi_\vu(\taus)\, \ee^{-t/\taur}$, we simply have $k\approx\taur^{-1}$ by \myeq{k-inv}, providing that $\taus\!\ll\!\taur$.
In such a case, we can simulate $\Phi_\vu(t)$ up to $\tstop\!>\!\taus$, and determine the reaction rate by a linear fitting of the long-time part in a logarithmic plot of $\Phi_\vu(t)$.
By fitting $\Phi_\vu$-curves of $Q\!=\!2,4$ in \myfig{fig3}{(a)} up to $\tstop\!=\!200$ (see \myfig{fig3}{(c)}), we obtain $\tau_2\!=\!727.7$ and $\tau_4\!=\!5407.7$ for the decaying time constants, which agree with $k^{-1}$ values of 736.9 and 5445.2 calculated using $\kV\!=\!0.315$ for $\ob^{-1}\!=\!5$ and $\kTST$ from \myeq{flux-over-pop} for $\oR^{-1}\!=\!5$.
We can roughly estimate the $\tstop$ dependence of sampling density by considering the ideal case of $\taus\!\rightarrow\!0$.
If we start $N_\mathrm{s}$ trajectories and let $N_\mathrm{r}$ be the number of trajectories that recross at $t\!>\!\taus$, then the sampling density is approximately constant for a short stop time, i.e.\ $N_\mathrm{r}/\tstop \propto N_\mathrm{s}$. However, with $\tstop$ decreasing, the average recrossing time decreases as $\tstop^2$, so for a given total simulation time, the sampling density will increase as $\tstop^{-2}$.
But in actual cases we can not arbitrarily decrease $\tstop$ due to other considerations.
%\begin{eqnarray*}
%    k^{-1} &=& -\int_{\taus}^\infty\! t \rmd \ee^{-t/\taur} \\
%    &=& -\left[ t\ee^{-t/\taur}|_{\taus}^\infty - \int_{\taus}^\infty\!\ee^{-t/\taur} \rmd t \right] \\
%    &=& (\taus + \taur) \ee^{-\taus/\taur}
%\end{eqnarray*}

In his famous paper~\cite{Zwanzig1973}, Zwanzig derived a generalized Langevin equation with $U(q)$ being the PMF for a reaction coordinate ($q$) from a model Hamiltonian with $q$ coupled bilinearly to a bath of harmonic oscillators ($\{x_i\}$).
By suitably choosing the coupling parameters, one can obtain the standard Langevin equation of \myeq{langevin} in the continuum limit.
In the vicinity of the barrier $q\!\sim\!0$, the dynamics of $\{q, x_i\}$ can be decoupled into an unstable normal mode ($\tilde{q}$) and other modes ($\{\tilde{x}_i\}$) through an orthogonal transformation.
Especially, the reaction coordinate $q$ is related to normal modes by $q = c_0 \tilde{q}+\sum c_i \tilde{x}_i$, where $c_0$ can be assumed positive.
Then according to Pollak's work~\cite{Pollak1986}, $\kK$ is simply $\jpos_{\tilde{q}}/\jpos_q$, where $\jpos_q$ and $\jpos_{\tilde{q}}$ are the fluxes through the dividing hyperplane of $q\teq 0$ and $\tilde{q}\teq 0$ at the saddle point, respectively.
Any trajectory associated with $\jpos_{\tilde{q}}$ will have $\tilde{q}\!\rightarrow\!\infty$, and is necessarily associated with $\kV \jpos_q$.
However, we have shown that $\kV\jpos_q\!>\!\kK \jpos_q\!=\!\jpos_{\tilde{q}}$, which indicates that there exist additional trajectories in $\kV\jpos_q$ that evolve into the hyperplane of $\tilde{q}\!=\!0$.
The above analysis demonstrates that the reaction coordinate determines the very problem of the reaction rate: $k\!\propto\!\kV\jpos_q$ for $q$, while $k\!\propto\!\jpos_{\tilde{q}}\!=\!\kK\jpos_q$ for $\tilde{q}$.
It will underestimate the reaction rate if $\jpos_{\tilde{q}}\!=\!\kK\jpos_q$ is used for the reaction coordinate $q$.

%Then according to Pollak's work~\cite{Pollak1986}, $\kK$ is simply $\jpos_{\tilde{q}}/\jpos_q$, where $\jpos_q$ is the flux through the dividing hyperplane of $q\teq 0$ at the saddle point, and similarly $\jpos_{\tilde{q}}$ is the flux for the unstable normal mode, $\tilde{q}=c_0 q+\sum c_i x_i$.
%Since there is no recrossing for $\tilde{q}(t)$, $\jpos_{\tilde{q}}$ has been considered as the actually reactive flux between the reactant and the product defined as minimums of $U(q)$.
%However, we have shown that $\kV\!>\!\kK$, which indicates that there exist trajectories that, after crossing $q\teq 0$ in the case of $Q\teq \infty$, neither return back to $q\!<\!0$ nor cross $\tilde{q}\teq 0$.
%The possibility of such trajectories is not inexplicable considering that $\tilde{q}$ depends on an infinite number of bath degrees of freedom in addition to $q$.
%The above analysis demonstrates that the definition of the dividing surface or, equivalently, the reaction coordinate determines the very problem of the reaction rate.

In summary, we have solved the recurrence time problem of Brownian motion on the barrier maximum, and given a rigorous interpretation of the transmission coefficient. We have proposed methods to determine reaction rates by short-time simulations from the barrier maximum, and discussed the relation to the reactive flux method and the significance of reaction coordinates for the reaction rate problem.

\begin{acknowledgments}
This work was supported by the Mid-career Researcher Program (NRF-2020R1A2C2101636), Bio \& Medical Technology Development Program (NRF-2022M3E5F3080873), Medical Research Center (MRC) grant (NRF-2018R1A5A2025286), and the Brain Pool Program (NRF-2021H1D3A2A02081370) funded by the Ministry of Science and ICT (MSIT) through the National Research Foundation of Korea (NRF).
It was also supported by the Ewha Womans University Research Grant of 2021.
We thank T.\ N.\ L.\ V\~{u} for helpful discussions.
\end{acknowledgments}

\appendix*
\renewcommand{\thefigure}{A\arabic{figure}}
\setcounter{figure}{0}
\section{Appendix}

\begin{figure}
\includegraphics[scale=1.2]{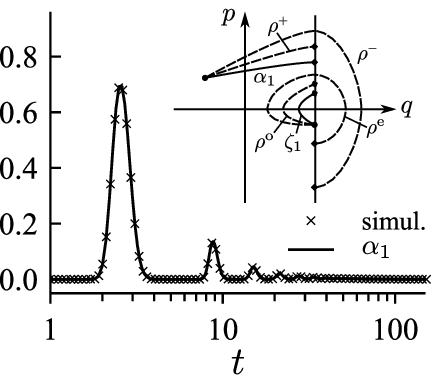}
\caption{\label{fig4} FPT distributions for the problem studied in Ref.~\cite{FPT2006}. The inset illustrates the meanings of involved quantities.}
\end{figure}

The winding property can also be used to solve the FPT problem of Brownian motion~\cite{Uhlenbeck1945}.
What we try to find is the probability density, $\alpha_1(t)$, that a particle starting from $(q_\mathrm{i},p_\mathrm{i})$ passes through $q\teq q_\mathrm{b}$ at time $t$ for the first time.
We assume $q_\mathrm{b}\tgt q_\mathrm{i}$, $p\tgt 0$ and $p_0\tgt 0$, and define $\gaussq^{\spm}(p,t)\equiv\gauss(q_\mathrm{b},\pm p,t;q_\mathrm{i},p_\mathrm{i})$ and $\gaussq^{\mathrm{o,e}}(p,t;p_0)\equiv\gauss(q_\mathrm{b},\pm p,t;q_\mathrm{b},-p_0)$ (see inset of \myfig{fig4}).
We can similarly determine the first-time probability density of winding around $(q_\mathrm{b},0)$, $\zeta_1(p,t;p_0)$, from the relation of $\zeta_1=\gaussq^\mathrm{o}-\zeta_1\odot\gaussq^\mathrm{e}$.
Then we can calculate the FPT distribution by
\[
    \alpha_1(t) = \int_0^\infty\!\rmd p\, p \left[\,\gaussq^{\splus}(p,t)-\zeta_1(p,t;p_0)\odot\gaussq^{\sminus}(p_0,t)\,\right].
\]
We test the approach by solving the problem presented in Sec.\ III of Ref.~\cite{FPT2006}.
Compared with the FPT distribution calculated up to three terms of the Rice series~\cite{Rice1945} (Fig.\ 2 of Ref.~\cite{FPT2006}), our result is well converged as shown in \myfig{fig4}.

%\bibliography{esc-barr}% Produces the bibliography via BibTeX.

\begin{thebibliography}{22}%
\makeatletter
\providecommand \@ifxundefined [1]{%
 \@ifx{#1\undefined}
}%
\providecommand \@ifnum [1]{%
 \ifnum #1\expandafter \@firstoftwo
 \else \expandafter \@secondoftwo
 \fi
}%
\providecommand \@ifx [1]{%
 \ifx #1\expandafter \@firstoftwo
 \else \expandafter \@secondoftwo
 \fi
}%
\providecommand \natexlab [1]{#1}%
\providecommand \enquote  [1]{``#1''}%
\providecommand \bibnamefont  [1]{#1}%
\providecommand \bibfnamefont [1]{#1}%
\providecommand \citenamefont [1]{#1}%
\providecommand \href@noop [0]{\@secondoftwo}%
\providecommand \href [0]{\begingroup \@sanitize@url \@href}%
\providecommand \@href[1]{\@@startlink{#1}\@@href}%
\providecommand \@@href[1]{\endgroup#1\@@endlink}%
\providecommand \@sanitize@url [0]{\catcode `\\12\catcode `\$12\catcode
  `\&12\catcode `\#12\catcode `\^12\catcode `\_12\catcode `\%12\relax}%
\providecommand \@@startlink[1]{}%
\providecommand \@@endlink[0]{}%
\providecommand \url  [0]{\begingroup\@sanitize@url \@url }%
\providecommand \@url [1]{\endgroup\@href {#1}{\urlprefix }}%
\providecommand \urlprefix  [0]{URL }%
\providecommand \Eprint [0]{\href }%
\providecommand \doibase [0]{https://doi.org/}%
\providecommand \selectlanguage [0]{\@gobble}%
\providecommand \bibinfo  [0]{\@secondoftwo}%
\providecommand \bibfield  [0]{\@secondoftwo}%
\providecommand \translation [1]{[#1]}%
\providecommand \BibitemOpen [0]{}%
\providecommand \bibitemStop [0]{}%
\providecommand \bibitemNoStop [0]{.\EOS\space}%
\providecommand \EOS [0]{\spacefactor3000\relax}%
\providecommand \BibitemShut  [1]{\csname bibitem#1\endcsname}%
\let\auto@bib@innerbib\@empty
%</preamble>
\bibitem [{\citenamefont {Kramers}(1940)}]{Kramers1940}%
  \BibitemOpen
  \bibfield  {author} {\bibinfo {author} {\bibfnamefont {H.~A.}\ \bibnamefont
  {Kramers}},\ }\href
  {https://www.sciencedirect.com/science/article/pii/S0031891440900982}
  {\bibfield  {journal} {\bibinfo  {journal} {Physica}\ }\textbf {\bibinfo
  {volume} {7}},\ \bibinfo {pages} {284} (\bibinfo {year} {1940})}\BibitemShut
  {NoStop}%
\bibitem [{\citenamefont {H\"anggi}\ \emph {et~al.}(1990)\citenamefont
  {H\"anggi}, \citenamefont {Talkner},\ and\ \citenamefont
  {Borkovec}}]{Hanggi1990}%
  \BibitemOpen
  \bibfield  {author} {\bibinfo {author} {\bibfnamefont {P.}~\bibnamefont
  {H\"anggi}}, \bibinfo {author} {\bibfnamefont {P.}~\bibnamefont {Talkner}},\
  and\ \bibinfo {author} {\bibfnamefont {M.}~\bibnamefont {Borkovec}},\ }\href
  {https://doi.org/10.1103/RevModPhys.62.251} {\bibfield  {journal} {\bibinfo
  {journal} {Rev. Mod. Phys.}\ }\textbf {\bibinfo {volume} {62}},\ \bibinfo
  {pages} {251} (\bibinfo {year} {1990})}\BibitemShut {NoStop}%
\bibitem [{\citenamefont {Pollak}\ and\ \citenamefont
  {Talkner}(2005)}]{Pollak2005}%
  \BibitemOpen
  \bibfield  {author} {\bibinfo {author} {\bibfnamefont {E.}~\bibnamefont
  {Pollak}}\ and\ \bibinfo {author} {\bibfnamefont {P.}~\bibnamefont
  {Talkner}},\ }\href {https://doi.org/10.1063/1.1858782} {\bibfield  {journal}
  {\bibinfo  {journal} {Chaos}\ }\textbf {\bibinfo {volume} {15}},\ \bibinfo
  {pages} {026116} (\bibinfo {year} {2005})}\BibitemShut {NoStop}%
\bibitem [{\citenamefont {Einstein}(1905)}]{Einstein1905}%
  \BibitemOpen
  \bibfield  {author} {\bibinfo {author} {\bibfnamefont {A.}~\bibnamefont
  {Einstein}},\ }\href {https://doi.org/10.1002/andp.19053220806} {\bibfield
  {journal} {\bibinfo  {journal} {Ann. Phys.}\ }\textbf {\bibinfo {volume}
  {322}},\ \bibinfo {pages} {549} (\bibinfo {year} {1905})}\BibitemShut
  {NoStop}%
\bibitem [{\citenamefont {Wigner}(1938)}]{Wigner1938}%
  \BibitemOpen
  \bibfield  {author} {\bibinfo {author} {\bibfnamefont {E.}~\bibnamefont
  {Wigner}},\ }\href {https://doi.org/10.1039/TF9383400029} {\bibfield
  {journal} {\bibinfo  {journal} {Trans. Faraday Soc.}\ }\textbf {\bibinfo
  {volume} {34}},\ \bibinfo {pages} {29} (\bibinfo {year} {1938})}\BibitemShut
  {NoStop}%
\bibitem [{\citenamefont {Wang}\ and\ \citenamefont
  {Uhlenbeck}(1945)}]{Uhlenbeck1945}%
  \BibitemOpen
  \bibfield  {author} {\bibinfo {author} {\bibfnamefont {M.~C.}\ \bibnamefont
  {Wang}}\ and\ \bibinfo {author} {\bibfnamefont {G.~E.}\ \bibnamefont
  {Uhlenbeck}},\ }\href {https://doi.org/10.1103/RevModPhys.17.323} {\bibfield
  {journal} {\bibinfo  {journal} {Rev. Mod. Phys.}\ }\textbf {\bibinfo {volume}
  {17}},\ \bibinfo {pages} {323} (\bibinfo {year} {1945})}\BibitemShut
  {NoStop}%
\bibitem [{\citenamefont {Blake}\ and\ \citenamefont
  {Lindsey}(1973)}]{Blake1973}%
  \BibitemOpen
  \bibfield  {author} {\bibinfo {author} {\bibfnamefont {I.}~\bibnamefont
  {Blake}}\ and\ \bibinfo {author} {\bibfnamefont {W.}~\bibnamefont
  {Lindsey}},\ }\href {https://doi.org/10.1109/TIT.1973.1055016} {\bibfield
  {journal} {\bibinfo  {journal} {IEEE Trans. Inf. Theory}\ }\textbf {\bibinfo
  {volume} {19}},\ \bibinfo {pages} {295} (\bibinfo {year} {1973})}\BibitemShut
  {NoStop}%
\bibitem [{\citenamefont {Eyring}(1935)}]{Eyring1935}%
  \BibitemOpen
  \bibfield  {author} {\bibinfo {author} {\bibfnamefont {H.}~\bibnamefont
  {Eyring}},\ }\href {https://doi.org/10.1063/1.1749604} {\bibfield  {journal}
  {\bibinfo  {journal} {J. Chem. Phys.}\ }\textbf {\bibinfo {volume} {3}},\
  \bibinfo {pages} {107} (\bibinfo {year} {1935})}\BibitemShut {NoStop}%
\bibitem [{\citenamefont {Evans}(1938)}]{Evans1938}%
  \BibitemOpen
  \bibfield  {author} {\bibinfo {author} {\bibfnamefont {M.~G.}\ \bibnamefont
  {Evans}},\ }\href {https://doi.org/10.1039/TF9383400049} {\bibfield
  {journal} {\bibinfo  {journal} {Trans. Faraday Soc.}\ }\textbf {\bibinfo
  {volume} {34}},\ \bibinfo {pages} {49} (\bibinfo {year} {1938})}\BibitemShut
  {NoStop}%
\bibitem [{\citenamefont {Farkas}(1927)}]{Farkas1927}%
  \BibitemOpen
  \bibfield  {author} {\bibinfo {author} {\bibfnamefont {L.}~\bibnamefont
  {Farkas}},\ }\href {https://doi.org/10.1515/zpch-1927-12513} {\bibfield
  {journal} {\bibinfo  {journal} {Z. Phys. Chem.}\ }\textbf {\bibinfo {volume}
  {125U}},\ \bibinfo {pages} {236} (\bibinfo {year} {1927})}\BibitemShut
  {NoStop}%
\bibitem [{\citenamefont {Pontryagin}\ \emph {et~al.}(1989)\citenamefont
  {Pontryagin}, \citenamefont {Andronov},\ and\ \citenamefont
  {Vitt}}]{Pontryagin1989}%
  \BibitemOpen
  \bibfield  {author} {\bibinfo {author} {\bibfnamefont {L.}~\bibnamefont
  {Pontryagin}}, \bibinfo {author} {\bibfnamefont {A.}~\bibnamefont
  {Andronov}},\ and\ \bibinfo {author} {\bibfnamefont {A.}~\bibnamefont
  {Vitt}},\ }\bibinfo {title} {Appendix: On the statistical treatment of
  dynamical systems},\ in\ \href {https://doi.org/10.1017/CBO9780511897818.012}
  {\emph {\bibinfo {booktitle} {Noise in Nonlinear Dynamical Systems}}},\
  Vol.~\bibinfo {volume} {1},\ \bibinfo {editor} {edited by\ \bibinfo {editor}
  {\bibfnamefont {F.}~\bibnamefont {Moss}}\ and\ \bibinfo {editor}
  {\bibfnamefont {P.~V.~E.}\ \bibnamefont {McClintock}}}\ (\bibinfo
  {publisher} {Cambridge University Press},\ \bibinfo {year} {1989})\ p.\
  \bibinfo {pages} {329–348}\BibitemShut {NoStop}%
\bibitem [{\citenamefont {Leimkuhler}\ and\ \citenamefont
  {Matthews}(2013)}]{Leimkuhler2013}%
  \BibitemOpen
  \bibfield  {author} {\bibinfo {author} {\bibfnamefont {B.}~\bibnamefont
  {Leimkuhler}}\ and\ \bibinfo {author} {\bibfnamefont {C.}~\bibnamefont
  {Matthews}},\ }\href {https://doi.org/10.1093/amrx/abs010} {\bibfield
  {journal} {\bibinfo  {journal} {Appl. Math. Res. Express}\ }\textbf {\bibinfo
  {volume} {2013}},\ \bibinfo {pages} {34} (\bibinfo {year}
  {2013})}\BibitemShut {NoStop}%
\bibitem [{\citenamefont {McKean~Jr.}(1962)}]{McKean1962}%
  \BibitemOpen
  \bibfield  {author} {\bibinfo {author} {\bibfnamefont {H.~P.}\ \bibnamefont
  {McKean~Jr.}},\ }\href {https://doi.org/10.1215/kjm/1250524936} {\bibfield
  {journal} {\bibinfo  {journal} {J. Math. Kyoto Univ.}\ }\textbf {\bibinfo
  {volume} {2}},\ \bibinfo {pages} {227} (\bibinfo {year} {1962})}\BibitemShut
  {NoStop}%
\bibitem [{\citenamefont {Risken}(1996)}]{Risken1996}%
  \BibitemOpen
  \bibfield  {author} {\bibinfo {author} {\bibfnamefont {H.}~\bibnamefont
  {Risken}},\ }\bibinfo {title} {Solutions of the {K}ramers {E}quation},\ in\
  \href {https://doi.org/10.1007/978-3-642-61544-3_10} {\emph {\bibinfo
  {booktitle} {The Fokker-Planck Equation: Methods of Solution and
  Applications}}}\ (\bibinfo  {publisher} {Springer Berlin Heidelberg},\
  \bibinfo {address} {Berlin, Heidelberg},\ \bibinfo {year} {1996})\ pp.\
  \bibinfo {pages} {238--240}\BibitemShut {NoStop}%
\bibitem [{\citenamefont {Wong}(1966)}]{Wong1966}%
  \BibitemOpen
  \bibfield  {author} {\bibinfo {author} {\bibfnamefont {E.}~\bibnamefont
  {Wong}},\ }\href {http://www.jstor.org/stable/2946238} {\bibfield  {journal}
  {\bibinfo  {journal} {SIAM J. Appl. Math.}\ }\textbf {\bibinfo {volume}
  {14}},\ \bibinfo {pages} {1246} (\bibinfo {year} {1966})},\ \bibinfo {note}
  {full publication date: Nov., 1966}\BibitemShut {NoStop}%
\bibitem [{\citenamefont {Wong}(1970)}]{Wong1970}%
  \BibitemOpen
  \bibfield  {author} {\bibinfo {author} {\bibfnamefont {E.}~\bibnamefont
  {Wong}},\ }\href {http://www.jstor.org/stable/2099359} {\bibfield  {journal}
  {\bibinfo  {journal} {SIAM J. Appl. Math.}\ }\textbf {\bibinfo {volume}
  {18}},\ \bibinfo {pages} {67} (\bibinfo {year} {1970})}\BibitemShut {NoStop}%
\bibitem [{\citenamefont {Yamamoto}(1960)}]{Yamamoto1960}%
  \BibitemOpen
  \bibfield  {author} {\bibinfo {author} {\bibfnamefont {T.}~\bibnamefont
  {Yamamoto}},\ }\href {https://doi.org/10.1063/1.1731099} {\bibfield
  {journal} {\bibinfo  {journal} {J. Chem. Phys.}\ }\textbf {\bibinfo {volume}
  {33}},\ \bibinfo {pages} {281} (\bibinfo {year} {1960})}\BibitemShut
  {NoStop}%
\bibitem [{\citenamefont {Chandler}(1978)}]{Chandler1978}%
  \BibitemOpen
  \bibfield  {author} {\bibinfo {author} {\bibfnamefont {D.}~\bibnamefont
  {Chandler}},\ }\href {https://doi.org/10.1063/1.436049} {\bibfield  {journal}
  {\bibinfo  {journal} {J. Chem. Phys.}\ }\textbf {\bibinfo {volume} {68}},\
  \bibinfo {pages} {2959} (\bibinfo {year} {1978})}\BibitemShut {NoStop}%
\bibitem [{\citenamefont {Zwanzig}(1973)}]{Zwanzig1973}%
  \BibitemOpen
  \bibfield  {author} {\bibinfo {author} {\bibfnamefont {R.}~\bibnamefont
  {Zwanzig}},\ }\href {https://doi.org/10.1007/BF01008729} {\bibfield
  {journal} {\bibinfo  {journal} {J. Stat. Phys.}\ }\textbf {\bibinfo {volume}
  {9}},\ \bibinfo {pages} {215} (\bibinfo {year} {1973})}\BibitemShut {NoStop}%
\bibitem [{\citenamefont {Pollak}(1986)}]{Pollak1986}%
  \BibitemOpen
  \bibfield  {author} {\bibinfo {author} {\bibfnamefont {E.}~\bibnamefont
  {Pollak}},\ }\href {https://doi.org/10.1063/1.451294} {\bibfield  {journal}
  {\bibinfo  {journal} {J. Chem. Phys.}\ }\textbf {\bibinfo {volume} {85}},\
  \bibinfo {pages} {865} (\bibinfo {year} {1986})}\BibitemShut {NoStop}%
\bibitem [{\citenamefont {Verechtchaguina}\ \emph {et~al.}(2006)\citenamefont
  {Verechtchaguina}, \citenamefont {Sokolov},\ and\ \citenamefont
  {Schimansky-Geier}}]{FPT2006}%
  \BibitemOpen
  \bibfield  {author} {\bibinfo {author} {\bibfnamefont {T.}~\bibnamefont
  {Verechtchaguina}}, \bibinfo {author} {\bibfnamefont {I.~M.}\ \bibnamefont
  {Sokolov}},\ and\ \bibinfo {author} {\bibfnamefont {L.}~\bibnamefont
  {Schimansky-Geier}},\ }\href {https://doi.org/10.1103/PhysRevE.73.031108}
  {\bibfield  {journal} {\bibinfo  {journal} {Phys. Rev. E}\ }\textbf {\bibinfo
  {volume} {73}},\ \bibinfo {pages} {031108} (\bibinfo {year}
  {2006})}\BibitemShut {NoStop}%
\bibitem [{\citenamefont {Rice}(1945)}]{Rice1945}%
  \BibitemOpen
  \bibfield  {author} {\bibinfo {author} {\bibfnamefont {S.~O.}\ \bibnamefont
  {Rice}},\ }\href {https://doi.org/10.1002/j.1538-7305.1945.tb00453.x}
  {\bibfield  {journal} {\bibinfo  {journal} {Bell Syst. Tech. J.}\ }\textbf
  {\bibinfo {volume} {24}},\ \bibinfo {pages} {46} (\bibinfo {year}
  {1945})}\BibitemShut {NoStop}%
\end{thebibliography}
%apsrev4-2.bst 2019-01-14 (MD) hand-edited version of apsrev4-1.bst
%Control: key (0)
%Control: author (8) initials jnrlst
%Control: editor formatted (1) identically to author
%Control: production of article title (-1) disabled
%Control: page (0) single
%Control: year (1) truncated
%Control: production of eprint (0) enabled
\providecommand{\noopsort}[1]{}\providecommand{\singleletter}[1]{#1}%

\end{document}